\documentclass[sigconf, screen, 9pt]{acmart}

\copyrightyear{2020}
\acmYear{2020}
\setcopyright{acmcopyright}\acmConference[NOSSDAV'20]{30th Workshop on
Network and Operating System Support for Digital Audio and Video}{June 10--11,
2020}{Istanbul, Turkey}
\acmBooktitle{30th Workshop on Network and Operating System Support for Digital
Audio and Video (NOSSDAV'20), June 10--11, 2020, Istanbul, Turkey}
\acmPrice{15.00}
\acmDOI{10.1145/3386290.3396930}
\acmISBN{978-1-4503-7945-8/20/06}

\usepackage{flushend,cuted}
\usepackage{graphicx}
\usepackage{subfigure}
\usepackage{booktabs} 
\usepackage{multirow}
\usepackage{euscript}
\usepackage{amssymb}
\usepackage{algorithm,algpseudocode}
\usepackage{enumitem}
\usepackage{colortbl}
\usepackage{framed}
\usepackage[misc]{ifsym}
\definecolor{mygray}{gray}{.9}
\setcopyright{acmcopyright}
\acmArticle{4}
\acmPrice{15.00}

\usepackage{dutchcal}
\begin{document}
\title{Self-play Reinforcement Learning for Video Transmission}
\author{Tianchi Huang$^{1}$, Rui-Xiao Zhang$^{1}$, Lifeng Sun$^{{1, 2, 3}*}$}
\affiliation{\emph{$^{1}$Dept. of CS \& Tech., $^{2}$BNRist, $^{3}$Key Laboratory of Pervasive Computing, Tsinghua University.}}

\renewcommand{\shortauthors}{Huang et al.}
\renewcommand{\authors}{Tianchi Huang, Rui-Xiao Zhang, Lifeng Sun}

\begin{abstract}
Video transmission services adopt adaptive algorithms to ensure users' demands.
Existing techniques are often optimized and evaluated by a function that linearly combines several weighted metrics.
Nevertheless, we observe that the given function fails to describe the requirement accurately. Thus, such proposed methods might eventually violate the original needs.
To eliminate this concern, we propose \emph{Zwei}, a self-play reinforcement learning algorithm for video transmission tasks. 
Zwei aims to update the policy by straightforwardly utilizing the actual requirement.
Technically, Zwei samples a number of trajectories from the same starting point, and instantly estimates the win rate w.r.t the competition outcome. 
Here the competition result represents which trajectory is closer to the assigned requirement. 
Subsequently, Zwei optimizes the strategy by maximizing the win rate.
To build Zwei, we develop simulation environments, design adequate neural network models, and invent training methods for dealing with different requirements on various video transmission scenarios. 
Trace-driven analysis over two representative tasks demonstrates that Zwei optimizes itself according to the assigned requirement faithfully, outperforming the state-of-the-art methods under all considered scenarios.
\end{abstract}




\maketitle

\section{Introduction}
Thanks to the dynamic growth of video encoding technologies and basic Internet services~\cite{cisco}, currently we are living with the great help of video transmission services. 
In particular, the videos are required to transmit with fulfilling users' requirements, where the requirement is often known as quality of experience~(QoE) or quality of service~(QoS).
Unfortunately, as much as the fundamental issue has already been published about \emph{two decades}~\cite{mo1999analysis}, current approaches, either heuristics or learning-based methods, fall short of achieving this goal. 
On the one hand, heuristics often use existing domain knowledge~\cite{mo1999analysis} as the working principle. However, such approaches sometimes require careful tuning and will backfire under the circumstance that violated with presumptions, finally resulting in the failure of achieving acceptable performance under all considered scenarios. On the other hand, learning-based methods~\cite{mao2017neural,zhang2019enhancing} leverage deep reinforcement learning~(DRL) to train a neural network~(NN) by interacting with the environments without any presumptions, aiming to obtain higher reward. In recent work, the reward function is often defined as a linear-based equation with the combination of manipulation variables. 
While in this study, we empirically discover that i) an inaccurate reward function may mislead the learning-based algorithm to generalize bad strategies, since ii) the actual requirement can hardly be presented by the linear-based reward function with fixed weights. Moreover, iii) considering the diversity of real-world network environments, it's difficult to present an accurate reward function that can perfectly fit \emph{any} network conditions~(\S\ref{sec:challenges}). 
As a result, despite its abilities to gain higher numerical reward score, learning-based schemes may generalize a strategy that hardly meet the actual requirements.

Taking a look from another perspective, we observe that the aforementioned problem can be naturally described as a deterministic goal or requirement~\cite{huang2018tiyuntsong}. 
For instance, in the most cases, the goal of the adaptive bitrate~(ABR) streaming algorithm is to achieve lower rebuffering time first, and next, reaching higher bitrate~\cite{Huang2019Hindsight, mao2019real}. 
Inspired by this opportunity, we envision a self-play reinforcement learning-based framework, known as Zwei, which can be viewed as a solution for tackling the video transmission dilemma~(\S\ref{sec:design}).
The key idea of Zwei is to sample trajectories repeatedly by itself, and distinguish the sequence that is closer to the assigned demand, so as to learn the strategy for satisfying the demand iteratively. 
Specifically, we apply Monte-Carlo~(MC) search to estimate moving decisions from the starting point. In the MC search process, several trajectories are sampled from the starting state according to the current policy. Then the expected long-term win rate is estimated by averaging the competition results from each trajectory pairs, where the result represents which one is closer to the actual demand between the two trajectories.
Having estimated the win rate, Zwei adopts the proximate policy optimization~(PPO)~\cite{schulman2017proximal} to optimize the NN via increasing the probabilities of the winning sample and reducing the possibilities of the failure sample.

In the rest of the paper~(\S\ref{sec:eval}), we attempt to evaluate the potential of Zwei using trace-driven analyses of various representative video transmission scenarios. To achieve this, we build several faithful video transmission simulators which can accurately replicate the environment via real-world network dataset. Specifically, we validate Zwei on two different tasks~(\S\ref{sec:background}), including client-to-server, and server-to-client service. Note that each of them has individual requirements and challenges. As expected, evaluation results demonstrate the superiority of Zwei against existing state-of-the-art approaches on all tasks. In detail, we show that \textbf{(i)} Zwei outperforms existing ABR algorithms on both HD videos and 4K videos, with the improvements on Elo score~\cite{coulom2008whole} of 32.24\% - 36.38\%. 
\textbf{(ii)} In the crowd-sourced live streaming~(CLS) scheduling task, Zwei reduces the overall costs on 22\% and decreases over 6.5\% on the overall stalling ratio in comparison of state-of-the-art learning-based scheduling method LTS~\cite{zhang2019enhancing}. 

\noindent \textbf{Contributions: }This paper makes four key contributions.
\begin{itemize}[leftmargin=*]
    \item We point out the shortcoming of learning-based schemes in video transmission tasks and present the idea that update networks without reward engineering~(\S\ref{sec:challenges}). 
    \item Zwei is a novel framework that aims to employ the self-play reinforcement learning method to make the idea practical~(\S\ref{sec:design}).
    \item We implement Zwei into two representative video transmission scenarios, i.e. rate adaptation and crowd-sourced scheduling. Results indicates that Zwei outperforms existing schemes on all considered scenarios~(\S\ref{sec:eval}).
\end{itemize}

\section{Background and Challenges}
\label{sec:back}

\subsection{Video Transmission Services}
\label{sec:background}
In this work, the video transmission service mainly consists of:

\textbf{\emph{Client-to-Server Service.}}~In this scenario, users often adopts a video player to watch the video on demand. First, video contents are pre-encoded and pre-chunked as several bitrate ladders on the server. Then the video player, placed on the client side, dynamically picks the proper bitrate for the next chunk to varying network conditions. Specifically, the bitrate decisions should achieve high bitrate and low rebuffering on the entire session~\cite{bentaleb2018survey,Huang2019Hindsight}. We call it \emph{adaptive bitrate streaming}~(ABR).
    
\textbf{\emph{Server-to-Client Service.}}~Considering that if we were the content provider and currently we had multiple content delivery networks~(CDNs) with different costs and performance, how to schedule the users' requests to the proper CDN, aiming to provide live streaming services withing less stalling ratio and lower cost? In common, we call that \emph{crowd-sourced live streaming(CLS)}~\cite{zhang2019enhancing}.
    

Details of each video transmission scenario are demonstrated in \S\ref{sec:eval}. We can see that the video transmission task is often required to obtain better performance under various mutual metrics.


\subsection{Challenges}
\label{sec:challenges}
\textbf{Motivation.}~Recent video transmission algorithms mainly consist of two types, i.e., heuristics and learning-based scheme. Heuristic methods utilize an existing fixed model or domain knowledge to construct the algorithm, while they inevitably fail to achieve optimal performance in all considered network conditions due to the inefficient parameter tuning~(\S\ref{related:heu},\cite{mao2017neural}). Learning-based schemes train a NN model towards the higher reward score from scratch, where the reward function is often defined as a linear combination of several weighted metrics~\cite{zhang2019enhancing, mao2017neural}. Nevertheless, considering the characteristics of the video transmission tasks above, \emph{we argue that the policy generated by the linear-based reward fails to always perform on the right track}. We, therefore, set up two experiments in the ABR scenario~(\ref{sec:abr}) to prove this conjecture. 

\begin{figure}
  \centering
    \includegraphics[width=0.48\linewidth]{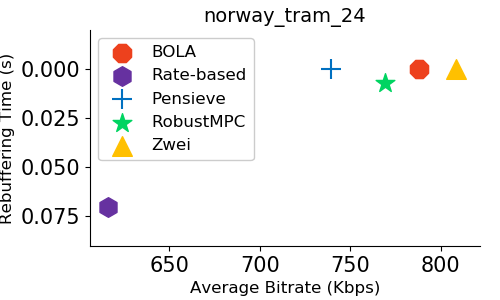}
    \includegraphics[width=0.48\linewidth]{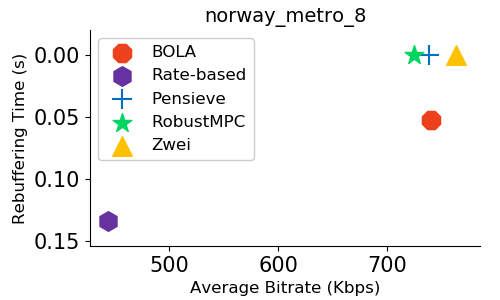}
  	\vspace{-5pt}
    \caption{We show average bitrate and rebuffering time for each ABR method. ABRs are performed over the HSDPA network traces.}
\label{fig:vmaf}
\end{figure}

\noindent $\textbf{Observation 1.}$ \emph{The best learning-based ABR algorithm Pensieve~\cite{mao2017neural} is not always the best scheme on every network traces.}

Recently, given a deterministic QoE metric with linearly combining of several underlying metrics, several attempts have been made to propose the ABR algorithm via model-based~\cite{yin2015control} or model-free reinforcement learning~(RL) method~\cite{mao2017neural}. 
However, such method heavily relies on the accuracy of the given QoE metric. Especially, \emph{how to set a proper QoE parameter for each network condition} is indeed a critical challenges for training a good ABR algorithm. In order to verify whether QoE parameters have influenced the performance of ABR algorithms, we set up an experiment to report average bitrate and rebuffering time of several state-of-the-art ABR baselines~(\S\ref{sec:abrbaseline}) and Zwei. Despite the outstanding performances that Pensieve achieves, the best learning-based ABR algorithm does not always stand for the best scheme. On the contrary, Zwei always outperforms existing approaches.

\begin{figure}
	\centering
  	\includegraphics[width=0.48\linewidth]{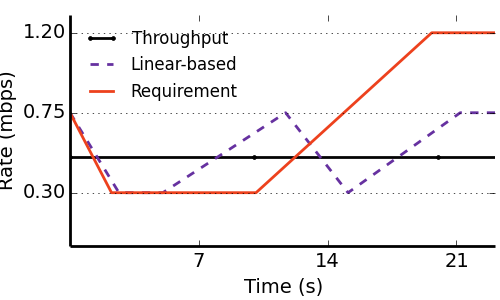}
  	\includegraphics[width=0.48\linewidth]{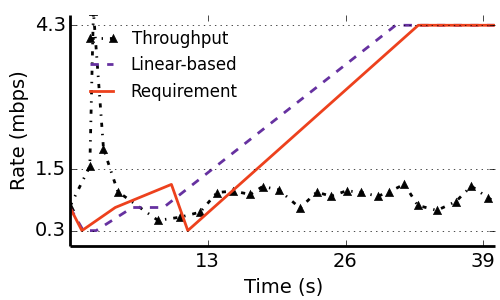}
  	\vspace{-5pt}
    \caption{The comparison of linear-based optimal and requirement-based optimal strategy. Results are evaluated on \emph{fixed~(0.5mbps)} and the real-world network trace~\cite{riiser2013commute}. We can see the difference between the actual requirement and the optimal trajectory generated by the linear-based reward.}
    \label{fig:case}
    \vspace{-15pt}
\end{figure}

\noindent $\textbf{Observation 2.}$ \emph{Recent linear-based weighted reward function can hardly map the actual requirement for all the network traces.}

To better understand the effectiveness of weighted-sum-based reward functions, we compare the optimal strategy of linear-based with requirement-based under two representative network traces, in which linear-based optimal is the policy which obtains maximum reward, and the requirement-based optimal stands for the closest strategy in terms of the given requirement. Unsurprisingly, from Figure~\ref{fig:case} we observe that linear-based optimal policy performs differently compared with the requirement-based optimal strategy. The reason is that the given weighted parameters are not allowed to be adjusted dynamically according to the current network status.
Generally, we believe that the policy learned by reward engineering might fall into the unexpected conditions. 


\noindent \textbf{Summary.} In general, we observe that no matter how precisely and carefully the parameter of the linear-based reward function tunes, such tuned functions can hardly meet the requirement of any network conditions. E.g., the parameter of stable and unstable network conditions are not the same.
To that end, traditional learning-based scheme, which often optimizes the NN via the assigned functions, will eventually fail to provide a reliable performance on any network traces. We therefore, attempt to learn the strategies from the original demands.

\section{Zwei Design}
\label{sec:design}
In this section, we briefly introduce the details of Zwei, including its key principle, training algorithms and implementation details. 


\begin{figure}
    \centering
  	\includegraphics[width=0.8\linewidth]{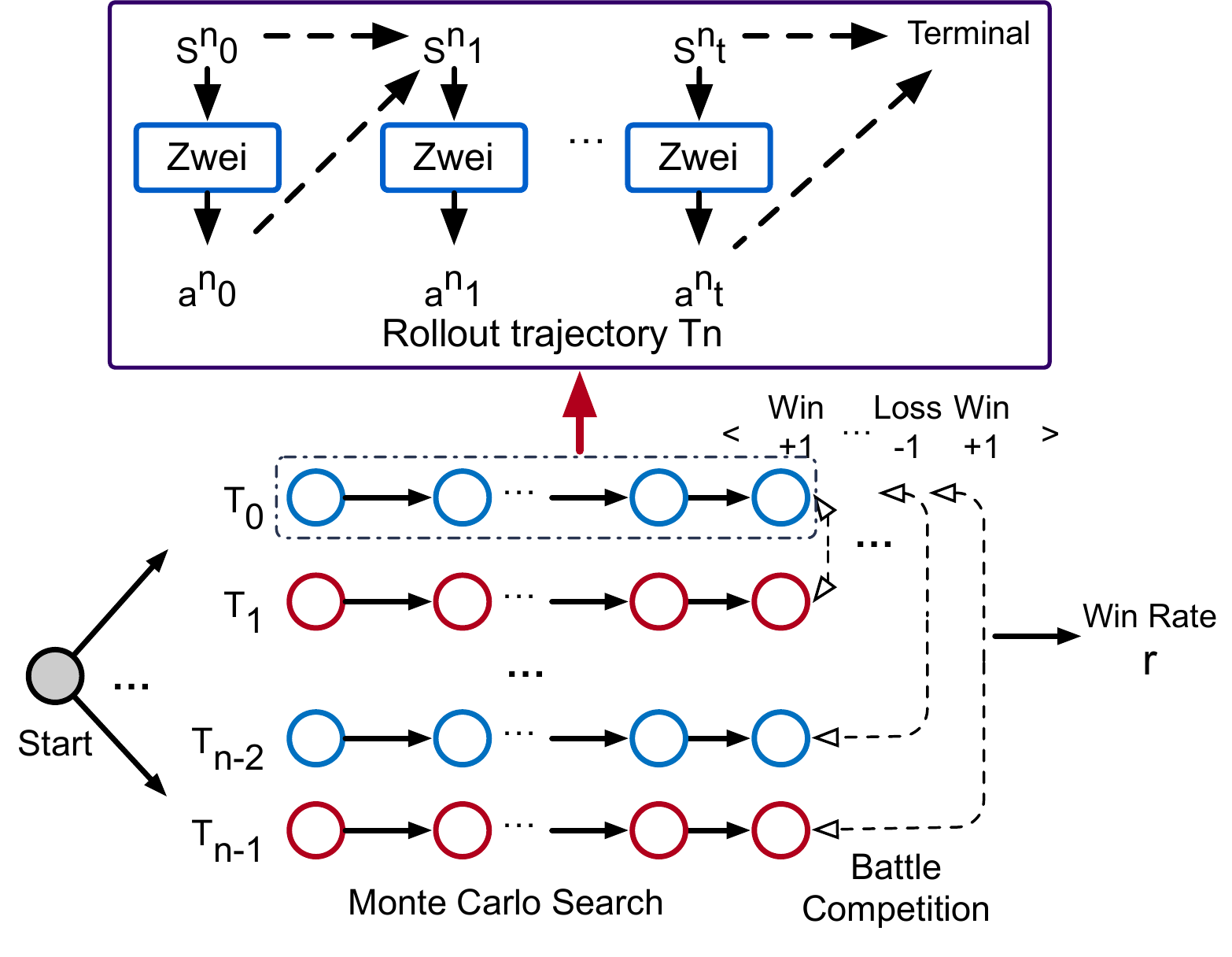}
  	\vspace{-10pt}
    \caption{Zwei System Overview. The framework is mainly composed of \emph{four} phases: Monte Carlo sampling, battle competition, win rate estimation, and policy optimization.}
    \label{fig:zweioverview}
\end{figure}

\subsection{Basic Idea}
\label{sec:selfplay}
As mentioned before, we attempt to generalize the strategy based on actual requirements instead of linear-based reward functions. Considering the basic requirement seldom directly provides gradients for optimization, we, therefore, employ the self-play method~\cite{silver2016mastering} that enables the NN to explore better policies and suitable rewards via self-learning.
Zwei treats the learning task as a competition between distinct trajectories sampled by itself, where the competition outcome is determined by a set of rules, symbolizing which one is closer to the given requirement. Subsequently, we are able to update the NN towards achieving a better outcome. 

\subsection{Training Methodology}
Figure~\ref{fig:zweioverview} presents the main phases in our framework. The pipeline can be summarized as follows: 

\textbf{Phase1: Monte Carlo Sampling.}~First, we adopt MC sampling method to sample $N$ different trajectories $T_{n}=\{s_{0}^n, a_{0}^n, s_{1}^n, a_{1}^n,$ $\dots, a_{t}^n\}, n \in N$ w.r.t the given policy $\pi(s)$ under the \emph{same environments}~($a_{t}\sim\pi(s_{t})$) and start point~(the gray point in Figure~\ref{fig:zweioverview}). Next, we record and analysis the underlying metric for the entire session. Finally, we store all the sample $T_{n}$ into $D$. Note that we can select Monte Carlo Tree Search~(MCTS)~\cite{silver2016mastering}, which is widely used in advanced research, to implement the process.

\textbf{Phase2: Battle Competition.}~To better estimate how the current policy performs, Zwei requires a module to label all trajectories from $D$: given two \emph{different} trajectories, $T_{i}$ and $T_{j}$ which are all collected from the \emph{same environment settings}~($T_{i}, T{j} \in D$)), we attempt to identify which trajectory is positive for NN updating, and which trajectory is generated by the worse policy. Thus, we implement a rule called $\texttt{Rule}$ which can determine the \emph{better} trajectory between the given two candidates, in which \emph{better} means which trajectory is closer to the requirement. At the end of the session, the terminal position $s_t$ is scored w.r.t the rules of the requirement for computing the game outcome $o$: $-1$ for a loss, $0$ for
a draw, and $+1$ for a win. The equation is listed in Eq.~\ref{eq:disc}. 
\begin{align}
    \label{eq:disc}
    o_{i}^{j} = \texttt{Rule}(T_{i}, T_{j}). \\
    s.t.\quad
    o_{i}^{j} = \{-1, 0, 1\}, T_{i}, T_{j} \in D, T_{i} \neq T_{j}.
\end{align}

\textbf{Phase3: Win Rate Estimation.}~Next, having computed the competition outcome $o_{i}$ for any two trajectories, we then attempt to estimate the average win rate $r_{i}$ for each trajectory $T_{i}$ in $D$, i.e., $r_{i} = \mathbb{E}[\texttt{Rule}(T_{i},)] = \lim_{N\to\infty}\frac{1}{N}\sum_{u}^{N} o_i^{u}$. Notice that the accuracy of the win rate estimation heavily depends on the number $N$ of trajectories. Since it's impractical to sample infinite number of samples in the real world, we further list the performance comparison of different sample numbers in \S\ref{sec:abr}.


\textbf{Phase4: Policy Optimization.}~
In this part, given a batch of collected samples and their win rate, our key idea is to update the policy via \emph{elevating} the probabilities of the winning sample from the collected trajectories and \emph{diminishing} the possibilities of the failure sample from the worse trajectories. In other words, the improved policy $\pi$ at state $s_t$ is required to pick the action $a_t$ which produced the best estimated win rate $r_t$.
We employ Proxy Policy Optimization (PPO)~\cite{schulman2017proximal}, a state-of-the-art actor-critic method, as to optimize the NN's policy. PPO uses clip method to restrict the step size of the policy iteration and update the NN by minimizing the following \emph{clipped surrogate objective}. We list the Zwei's loss function $\mathcal{L}^{\text{Zwei}}(\theta)$ in Eq.~(\ref{eq:adzwei}).
The function consists of a policy loss and a outcome loss. The policy are computed as Eq.~\ref{eq:policy}, 
where $p_t(\theta)$ denote the probability ratio between the policy $\pi_{\theta}$ and the old policy $\pi_{\theta_{\text{old}}}$, i.e.,  $\frac{\pi_{\theta}(a_t|s_t)}{\pi_{\theta_{\text{old}}}(a_t|s_t)}$; $ \epsilon$ is the hyper-parameters which controls the clip range. We set $\epsilon=0.2$ which consistent with the original paper~\cite{schulman2017proximal}. $\hat{A}_t$ is the advantage function~(Eq.~\ref{eq:advantage}),
$V_{\theta_{p}}(s)$ is provided by another value NN. 
Meanwhile, the value loss aims to minimize the mean square error between $r_t$ and $V_{\theta_{p}}(s)$~(Eq.~\ref{eq:value}). 
Moreover, we also add entropy $H(s_t; \theta)$ to encourage exploration. For more details, please refer to our repository~\cite{zweigit}. 
\begin{small}
\begin{align}
    \mathcal{L}^{\emph{Policy}}
    & = \hat{\mathbb{E}_{t}}\Big[ \min\Big( p_t(\theta)\hat{A}_t, \text{clip}\big(p_t(\theta), 1 - \epsilon, 1 + \epsilon\big)\hat{A}_t \Big) \Big]. 
    \label{eq:policy}
\end{align}
\begin{equation}
    \hat{A}_t = r_t - V_{\theta_{p}}(s_t)
    \label{eq:advantage}
\end{equation}

\begin{equation}
    \mathcal{L}^{\emph{Value}} = \frac{1}{2} \hat{\mathbb{E}_{t}}\left[ V_{\theta_{p}}(s_t) - r_t \right]^{2}.
    \label{eq:value}
\end{equation}

\begin{equation}
    \nabla \mathcal{L}^{Zwei} = \nabla_\theta \mathcal{L}^{\emph{Policy}} + \nabla_{\theta_{p}} \mathcal{L}^{\emph{Value}} + \nabla_\theta \beta H(s_t; \theta).
    \label{eq:adzwei}
\end{equation}
\end{small}

\section{Evaluation}
\label{sec:eval}
In this section, we thoroughly evaluate the performance of Zwei over two representative scenarios.

\subsection{Zwei's NN Implementation}
We use TensorFlow~\cite{abadi2016tensorflow} to construct Zwei. 
Zwei's policy network takes a n-dims vector with \emph{softmax} active function as the output, and the outcome network outputs a value with \emph{tanh} function scaled in $(-1, 1)$.
Considering the characteristics on video transmission tasks, we construct different NN architectures for each task, but use the same set of hyper-parameters for training the NN: sample number $N=16$, learning rate $\alpha=10^{-4}$. 
Notice that we discuss the NN architecture and the effect of different sample number in \S\ref{sec:abr}.

\subsection{ABR Scenarios}
\label{sec:abr}


\subsubsection{ABR's Background}
The ABR video streaming architecture consists of a video player client with a constrained buffer length and an HTTP-Server or Content Delivery Network~(CDN). The video player client decodes and renders video frames from the playback buffer. Once the streaming service starts, the client fetches the video chunk from the HTTP Server or CDN orderly by an ABR algorithm. The algorithm determines next chunks' video quality.
After finished to play the video, several metrics, such as total bitrate, total re-buffering time and total bitrate change will be summarized as a QoE metric to evaluate the performance.



\begin{figure*}
    \centering
    \subfigure[Zwei vs. ABR schemes]{
        \includegraphics[width=0.23\linewidth]{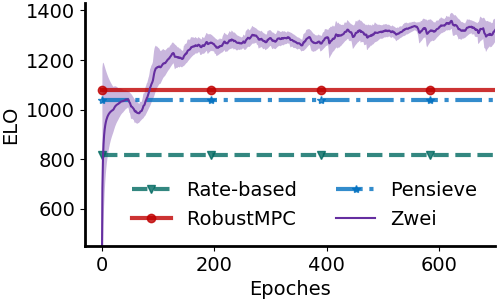}
        \label{fig:abr:a}}
    \subfigure[Zwei vs. Tiyuntsong]{
        \includegraphics[width=0.23\linewidth]{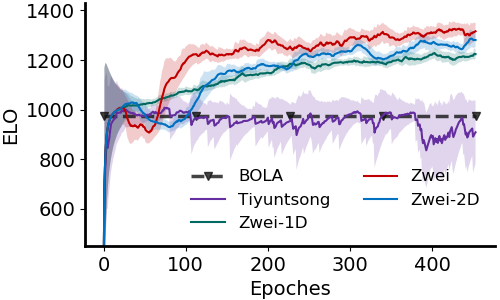}
        \label{fig:abr:b}}
    \subfigure[ABR Scheme Details]{
    \includegraphics[width=0.23\linewidth]{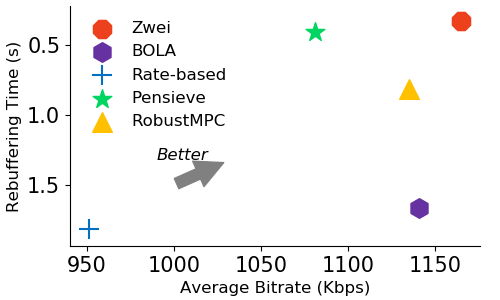}
        \label{fig:abr:c}}
    \subfigure[CDF of ABR Algorithms]{
    \includegraphics[width=0.23\linewidth]{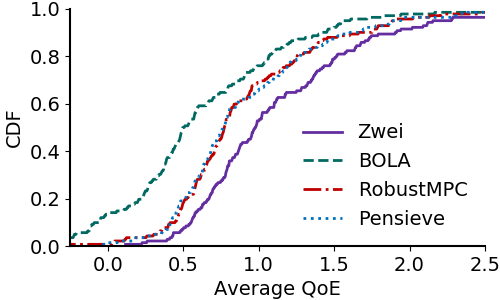}
        \label{fig:abr:d}}
    \vspace{-15pt}  
    \caption{This group of figures show the comparison results of Zwei and other ABR approaches. Results are evaluated in the typical ABR system with HD Videos~(video resolution=1920 $\times$ 1080, maximum bitrate=4.3mbps).}
    \label{fig:abr}
\end{figure*}

\subsubsection{NN Architecture}
We now explain the details of the Zwei's neural network~(NN), including its inputs, outputs, network architecture, and implementation.

\textbf{Inputs.}
For each chunk $k=8$, Zwei takes the state inputs $S_k=\{X_k, \tau_k, N_k, b_k, c_k, l_k\}$ as the input, in which $X_k$ represents the throughput measured for past $t$ times, $\tau_k$ reflects the vector of download time for past $t$ chunks, $N_k$ is the next video sizes for each bitrate chunk, $b_k$ means current buffer occupancy, $l_k$ is the normalized value of the last bitrate selected, $c_k$ is the scalar which means the video chunk remaining. 

\textbf{Outputs.}
We attempt to use discrete action space to describe the output. Note that the output is an n-dim vector indicating the probability of the bitrate being selected under the current ABR state $S_k$. In this work, we set $n=6$, which is widely used in ABR papers~\cite{huang2018tiyuntsong,mao2017neural}.

\textbf{NN Representation.}
Zwei's NN representation is quite simple: it leverages three fully-connected layers, which is sized 128, 64, and 64 respectively, for describing the feature extraction layer. The output of the NN's policy network is a \emph{6-dims} vector, which represents the probabilities for each bitrate selected. The NN's outcome network outputs a value scaled in (-1, 1).

\textbf{Requirements for ABR tasks.}
Algorithm~\ref{alg:Overall} is an example of Rule $\texttt{Rule}$ which used in the ABR scenario, where $\epsilon_{c}$ is a small number that can add noise for improving Zwei's robustness. In this work, we set $\epsilon_{c}$ as 10. Note that, such settings are trivial and you can just set the value as a small value~(e.g., 0.1) instead.

\begin{small}
\begin{algorithm}
\caption{Rule for the ABR task.} 
\label{alg:Overall} 
\begin{algorithmic}[1]
\Require Trajectory $T_{u}, T_{v}$;
\State Compute average bitrate $r_{u}, r_{v}$, average rebuffering $b_{u}, b_{v}$ from the given trajectories $T_{u}, T_{v}$. \Comment Requirements: i) low rebuffering time ii) high bitrate.
\State Initialize Return $s=\{-1, -1\}$;
\If{$\left| b_{u} - b_{v} \right|< \epsilon_{c}$ or $\left| r_{u} - r_{v} \right|< \epsilon_{c}$}
\State Randomly set $s_{0}$ or $s_{1}$ as 1. 
\Comment Add noise to improve the robustness.
\ElsIf{$\left| b_{u} - b_{v} \right|<\epsilon_{c}$}
\State $s_i \gets 1, i={argmax}_{i\in \{u,v\}}{r}$;
\Else
\State $s_i \gets 1, i={argmin}_{i\in \{u,v\}}{b}$;
\EndIf
\State return $s$;
\end{algorithmic} 
\end{algorithm}
\end{small}

\textbf{QoE Representation.}
Recall that the goal of ABR algorithm is to select bitrates for the next chunk $k$ with high bitrate $r_k$ and less rebuffering time $t_k$~\cite{mao2019real}, thus the optimization reward $q_k$ can be normally written as $q_k = r_k - \alpha t_k$, here $\alpha$ is the coefficient that adjust the importance of the underlying metrics. Note that prior research often add additional smoothness metric $\left| r_k - r_{k-1} \right|$ to control the bitrate change of the entire session, while in practice the following metric is neglectable for the ABR algorithm~\cite{mao2019real,Huang2019Hindsight}. Hence, in this work we also set the smoothness to zero for better understanding the fundamental performance of Zwei. Solving Zwei with smoothness metric will be our future work.

\begin{figure*}
    \centering
    \subfigure[Zwei vs. ABR schemes]{
        \includegraphics[width=0.23\linewidth]{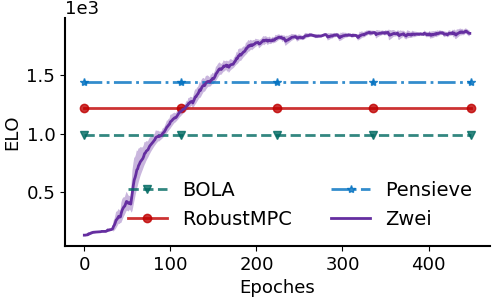}
        \label{fig:abr2:a}}
    \subfigure[Zwei with Different Samples]{
        \includegraphics[width=0.23\linewidth]{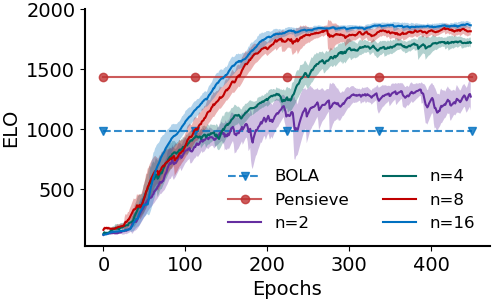}
        \label{fig:abr2:b}}
    \subfigure[ABR Algorithm Details]{
    \includegraphics[width=0.23\linewidth]{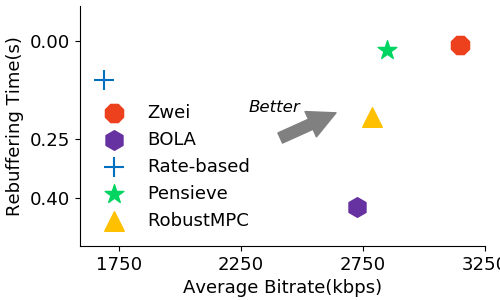}
    \label{fig:abr2:c}}
    \subfigure[CDF of ABR Algorithms]{
    \includegraphics[width=0.23\linewidth]{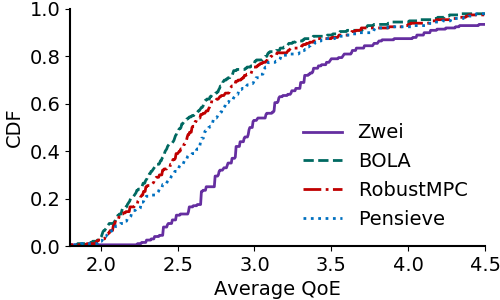}
    \label{fig:abr2:d}}
    \vspace{-15pt}
    \caption{This group of figures show the comparison results of Zwei and other ABR approaches. Results are evaluated in the typical ABR system with 4K Videos~(video resolution=3840 $\times$ 2160, maximum bitrate=12.0mbps).}
    \label{fig:abr2}
\end{figure*}

\subsubsection{Experimental Setup} We employ the standard ABR's emulation environment~\cite{mao2017neural} to evaluate Zwei. We adopt various network bandwidth databases, including HSDPA~\cite{riiser2013commute}, FCC~\cite{bworld}. Training process lasts approximate 45000 steps, within 10 hours to obtain a reliable result. In this experiment, we setup two ABR scenarios, i.e., HD and 4K video scenario. In the HD video scenario, we adopt \emph{EnvivioDash3}, a video that commonly used in recent work~\cite{mao2017neural}, to validate Zwei, where the video chunks are encoded as $\{0.3, 0.75, 1.2, 1.8, 2.8, 4.3\}$ mbps. In the 4K-video scenario, we use the popular open-source movie Big Buck Bunny~\cite{bbb2020}, which is now even more a world standard for video standards. Specifically, we pick 6 bitrates from \texttt{Dash.js} standard~\cite{dash}, i.e., $\{0.2, 0.6, 1.5, 4.0, 8.0, 12.0\}$ mbps.

\textbf{ABR Baselines.}
In this paper, we select several representational ABR algorithms from various type of fundamental principles:
\begin{enumerate}[leftmargin=*]
    \label{sec:abrbaseline}
    \item \textbf{\emph{Rate-based}~\cite{jiang2014improving}:}~the basic baseline for ABR problems. It leverages \emph{harmonic mean} of past five throughput measured as future bandwidth.
    
    \item \textbf{\emph{BOLA}~\cite{spiteri2016bola}:}~the most popular buffer-based ABR scheme in practice. BOLA turns the ABR problem into a utility maximization problem and solve it by using the Lyapunov function.
    
    \item \textbf{\emph{RobustMPC}~\cite{yin2015control}:}~a state-of-the-art heuristic method which maximizes the objectives by jointly considered the buffer occupancy and throughput predictions. We implement \emph{RobustMPC} by ourselves.

    \item \textbf{\emph{Pensieve}~\cite{mao2017neural}:}~the state-of-the-art learning-based ABR scheme which utilizes DRL to select bitrate for next video chunks.
    
    \item \textbf{\emph{Tiyuntsong}~\cite{huang2018tiyuntsong}:}~the first study of multi-objective optimization ABR approach. Tiyuntsong uses actor-critic method to update the NN via the competition with two agents under the same network condition. 
\end{enumerate}

\textbf{Evaluation Metrics} The Elo rating~\cite{coulom2008whole} is a traditional method for calculating the relative performance of players in zero-sum games. It's suitable to compare different schemes via win rate information only. Thus, we also use Elo score to compare different ABR schemes, similar to that of AlphaGo~\cite{silver2016mastering} and Tiyuntsong~\cite{huang2018tiyuntsong}.


\subsubsection{HD Video} We study the performance of Zwei over the HD video dataset and HSDPA network traces. Followed by recent work~\cite{yin2015control}, we set $\alpha$ = 4.3 as the basic QoE-HD metric.

\textbf{Zwei vs. Existing schemes.}~
In this experiment, we compare Zwei with existing ABR schemes over the HSDPA dataset. Results are computed as the Elo-score and reported in Figure~\ref{fig:abr:a}. Specifically, we first select several previously proposed approaches and validate the performance respectively under the \emph{same} network and video environment. Next, we use \texttt{Rule} to estimate their winning rate. Finally, we compute the Elo rating for these approaches.

Through the experiments we can see that Zwei outperforms recent ABR approaches. In particular, Zwei improves Elo-score on 36.38\% compared with state-of-the-art learning-based method Pensieve, and increases 31.11\% in terms of state-of-the-art heuristic method RobustMPC. Same conclusion are demonstrated as the CDF of QoE in Figure~\ref{fig:abr:d}.
Besides, we also illustrate the comparison results of different methods on average bitrate and average rebuffering in Figure~\ref{fig:abr:c}. As shown, Zwei can not only achieve highest bitrate but also obtain lowest rebuffering under all network traces. 
We compare Zwei with previously proposed self-learning method Tiyuntsong on the same experimental setting. Results are shown as the Elo-curve in Figure~\ref{fig:abr:b}. As expected, Zwei with any NN architectures outperforms Tiyuntsong on average Elo-score of 35\%.
The reason is that Tiyuntsong treat the learning process as a battle with two separated NNs, as each agent have to use the sample only collected by the agent itself. Thus, it lacks sample efficiency and sometimes struggles into the sub-optimal solution. 

\textbf{Zwei with Different NN Architectures.}
This experiment considers Zwei with several NN architectures, where Zwei-1D is the standard ABR NN architecture~\cite{mao2017neural}, Zwei-2D uses stacked Conv-2D layers to extract features, and Zwei takes three fully-connected layers sized $\{128, 64, 64\}$. Results are shown in Figure~\ref{fig:abr:b}.  Unsurprisingly, when Zwei trains with some complicated NN architecture~(i.e., Zwei-1D and Zwei-2D), it performs worse than the fully connected NN scheme. This makes sense since ABR is a light-weighted task which can be solved in a practical and uncomplicated manner instead of a NN incorporating some \emph{deep} yet \emph{wide} layers.



\subsubsection{4K Video} we evaluate Zwei with 4K videos on HDFS network traces, and the video are encoded with $\{0.2, 0.6, 1.5, 4.0, 8.0, 12.0\}$ mbps.
In particular, we have also retrained Pensieve with the Big Buck Bunny and QoE-4K. 

\textbf{QoE metrics for other approaches.} Due to the difference between the maximum bitrate of 4K video~(i.e., 12mbps) and HD video~(i.e., 4.3mbps), we refer $QoE_4K$ as the QoE function for other ABRs~(expect Zwei). Specifically, we set the penalty of rebuffering parameter $\alpha$ to 20 for better avoiding the occurrence of rebuffering.

\textbf{Zwei vs. Recent ABR Schemes.} Figure~\ref{fig:abr2:a} plots the learning curves in terms of the comparison of Zwei and other ABR methods. We observe that Zwei has already achieved state-of-the-art performance in almost 200 epochs, and finally,  the performance improvement of Zwei can achieve 32.24\% against the second best algorithm Pensieve and 58.58\%-120.76\% against others. What's more, as depicted in Figure~\ref{fig:abr2:c}, no matter average bitrate or total rebuffering time, Zwei always stands for the best scheme among all candidates. In particular, Zwei increases 10.46\% on average bitrates with reducing 56.52\% on total rebuffering time. Meanwhile, we also report the comparison of QoE performance of the proposed scheme in Figure~\ref{fig:abr2:d}. As expected, Zwei outperforms existing ABR algorithms without reward engineering, with the increasing on average QoE of 11.59\%.

\textbf{Zwei with Different Samples $N$.}~Besides, we also investigate the Elo-rating of Zwei with different sample $N$, where we set $N=\{2,4,8,16\}$. Figure~\ref{fig:abr2:b} illustrates the comparison of each method, we observe that Zwei can achieve sample efficiency with the increasing of the sample number $N$.

\subsection{CLS Scheduling}
\label{sec:cls}


\begin{figure}
    \centering
    \subfigure[Zwei vs. Existing Methods]{
        \includegraphics[width=0.48\linewidth]{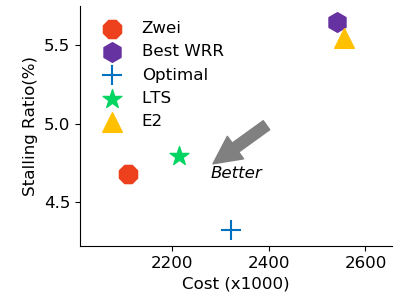}
        \label{fig:clszwei}
    }\subfigure[Zwei Training Curve]{
        \includegraphics[width=0.48\linewidth]{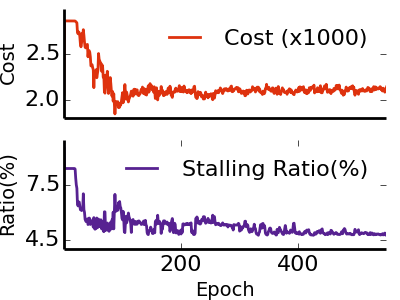}
        \label{fig:clstraining}
    }
    \vspace{-15pt}
    \caption{Results of Zwei under the CLS environment.}
    \vspace{-10pt}
    \label{fig:lts2}
\end{figure}

\subsubsection{CLS Overview}
The ClS system is composed of a source server and several CDNs. Upon received viewers' requests, the CLS platform will first aggregate all stream data to the source server, and then deliver the video stream to viewers through CDN providers according to a certain scheduling strategy.
\vspace{-5pt}
\subsubsection{Implementation}
Our experiments are conducted on the real-world CLS dataset provided by the authors~\cite{zhang2019enhancing}, spanning 1 week (6 days for training and 1 day for test). At each time, we select 3 candidates from total 4 different CDN providers, and we fit a separate simulator for them. Followed by previous work~\cite{zhang2019enhancing}, we use a piecewise linear model to characterize this relationship between workload and block ratio. Note that the following features are extracted by the real CDN dataset. What's more, we set the CDN pricing model w.r.t various CDN providers in industry, such as Amazon E2 and Tencent CDN. See~\cite{zweigit} for details.
\vspace{-5pt}

\begin{small}
\begin{algorithm}
\caption{Rule for the CLS task.} 
\label{alg:Overallcls} 
\begin{algorithmic}[1]
\Require Trajectory $T_{u}, T_{v}$;
\State Compute average stalling ratio ${stall}_{u}, {stall}_{v}$, accumulative cost $c_{u}, c_{v}$ from the given trajectories $T_{u}, T_{v}$. \Comment Requirements: i) low stalling ratio ii) low costs.
\State Initialize Return $s=\{-1, -1\}$;
\If{$\left| {stall}_{u} - {stall}_{v} \right|< \epsilon$ or $\left| c_{u} - c_{v} \right|< \epsilon_{c}$}
\State Randomly set $s_{0}$ or $s_{1}$ as 1. 
\Comment Add noise to improve the robustness.
\ElsIf{$\left| {stall}_{u} - {stall}_{v} \right|<\epsilon_{c}$}
\State $s_i \gets 1, i={argmin}_{i\in \{u,v\}}{c}$;
\Else
\State $s_i \gets 1, i={argmin}_{i\in \{u,v\}}{stall}$;
\EndIf
\State return $s$;
\end{algorithmic} 
\end{algorithm}
\end{small}
\vspace{-10pt}
\textbf{Baselines.}
Like other scenarios, we also compare Zwei with the following state-of-art scheduling baselines:
\begin{enumerate}[leftmargin=*]
    \item \textbf{\emph{Weighted round robin~(WRR)}~\cite{adhikari2012unreeling}.} The requests will be redirected to different CDN providers w.r.t a constant ratio. We adopt the algorithm with the \emph{best} parameters.

    \item \textbf{\emph{E2}~\cite{jiang2017pytheas}.}~Exploitation and Exploration~(E2) algorithm utilizes harmonic mean for estimating CDN providers' performance, and select with the highest upper confidence bound of reward. We use E2 algorithm provided by the authors~\cite{jiang2017pytheas}.
    
    \item \textbf{\emph{LTS}~\cite{zhang2019enhancing}.}~State-of-the-art CLS algorithm which uses deep reinforcement learning to train the NN towards lower stalling ratio. However, it ignores the trade-off between the cost and the performance. We evaluate LTS by Zhang et. al.~\cite{zhang2019enhancing}.
\end{enumerate}

\textbf{NN Representation.}
We implement Zwei in CLS as suggested by recent work~\cite{zhang2019enhancing}. More precisely speaking, for each CDN provider $i$, Zwei passes past 20 normalized workload and stalling ratio into Conv-2D layers with filter=64, size=4, and stride=1. Then several output layers are merged into a hidden layer that uses 64 units to apply the \emph{softmax} activation function. We model the action space as a heuristic way: each CDN provider has 3 choices, i.e., incrementally increases its configuration ratio by 1\%, 5\%, and 10\%. 

\textbf{Requirements for CLS tasks.}
Algorithm~\ref{alg:Overallcls} describes the $\texttt{Rule}$ of CLS. Note that the $\epsilon_{c}=10$ is also a small number that can add noise to improve Zwei's robustness.

\textbf{Zwei vs. State-of-the-art Scheduling Methods}
We start to study how well that Zwei achieves under the CLS scenario. As shown in Figure~\ref{fig:clszwei}, we find that Zwei stands for the best scheme among the candidates. Specifically, Zwei reduces the overall costs on 22\% compared with state-of-the-art learning-based method LTS, and decreases over 6.5\% in terms of the overall stalling ratio. The reason is that LTS takes the weight-sumed combination function as the reward, while the function can hardly give a clearer guidance for the optimized algorithm. Moreover, comparing the performance of Zwei with the optimal strategy generated by the reward function, we observe that both optimal policy and Zwei are in the Pareto front. Zwei consumes less pricing costs than the optimal policy since the requirement is to \emph{minimize the cost first}.

\textbf{Details.}
Besides, we also present the training process in Figure~\ref{fig:clstraining}. As shown, Zwei converges in less than 600 epochs, which needs about 3 hours. It's worth noting that Zwei also experienced two stages on the CLS task. The first stage ranges from 0 to 100 epochs, and we can see the goal of Zwei is to minimize the cost without considering the number of stalling ratio. The rest of the process we find that Zwei attempts to reduce the number of stalling ratio. Meanwhile, the cost curve converges to a steady state. 

\vspace{-10pt}
\section{Related Work}
\textbf{Heuristic Methods.}
\label{related:heu}
Heuristic-based ABR methods often adopts throughput prediction~(E.g., FESTIVE~\cite{jiang2014improving}) or buffer occupancy control~(E.g., BOLA~\cite{spiteri2016bola}) to handle the task.
However, such approaches suffer from either inaccurate bandwidth estimation or long-term bandwidth fluctuation problems.
Then, MPC~\cite{yin2015control} picks next chunks' bitrate by jointly considering throughput and buffer occupancy. Nevertheless, MPC is sensitive to its parameters since the it relies on well-understanding different network conditions. 

Through preliminary measurements, it is widely accepted that the strategies are largely statically configured~\cite{adhikari2012unreeling}. In recent years, dynamic scheduling across different CSPs has received more attention. Jiang et. al.~\cite{jiang2017pytheas} uses E2 method to replace traditional model-based methods. 

\textbf{Learning-based Schemes.}
Recent years, several learning-based attempts have been made to tackle video transmission problem. For example, Mao et al.~\cite{mao2017neural} develop an ABR algorithm that uses DRL to select next chunks' bitrate.
Tiyuntsong optimizes itself towards a rule or a specific reward via the competition with two agents under the same network condition~\cite{huang2018tiyuntsong}. 
LTS~\cite{zhang2019enhancing} is a DRL-based scheduling approach which outperforms previously proposed CLS approaches. 
However, such methods fail to achieve actual requirements since they are optimized via a linear-based reward function.
\vspace{-10pt}
\section{Conclusion}
We propose Zwei, which utilize self-play RL to enhance itself based on the actual requirement, where the requirement is always hard to be described as a linear-based manner. We show that Zwei outperforms recent work with the improvements of more than 22\% on two representative video transmission tasks.
~(32.24\% on Elo score over the ABR scenario, 22\% on stalling ratio over the LTS).

\noindent \textbf{Acknowledgement.}~We thank the \emph{MMSys} and \emph{NOSSDAV} reviewer for the valuable feedback. This work was supported by NSFC under Grant 61936011, 61521002, Beijing Key Lab of Networked Multimedia, and National Key R\&D Program of China (No. 2018YFB1003703).

\bibliographystyle{ACM-Reference-Format}
\bibliography{ref}
\appendix
\section{Appendix}
This supplementary material details the principle of Zwei. Due to the length of the supplemental material, we list a content to facilitate the selection of interested parts for review. Although these contents have \textbf{NOT} appeared in the main text, we believe that they will help the reviewer get a \emph{thorough understanding} of Zwei.

\subsection{Additional Experiments for ABR Tasks}
We list the experimental details under validation set from Table~\ref{detail:1} to Table~\ref{detail:3}, where the ABR algorithm includes Zwei, BOLA, RobustMPC, Pensieve and Rate-based. The validation set contains several different network scenarios. E.g., \emph{bus-1} represents network trace No.1 which collected from the bus.  
\begin{table*}[]
    \centering
    \caption{Details of ABR Tasks}
    \label{detail:1}
    \begin{tabular}{c|c|c|c|c|c|c|c|c|c|c}
    \hline
        Trace Name & \multicolumn{2}{|c|}{BOLA} & \multicolumn{2}{|c|}{Rate-based} & \multicolumn{2}{|c|}{Pensieve} & \multicolumn{2}{|c|}{RobustMPC} & \multicolumn{2}{|c}{Zwei} \\ \hline
        - & Avg. Bit. & Rebuf. & Avg. Bit. & Rebuf. & Avg. Bit. & Rebuf. & Avg. Bit. & Rebuf. & Avg. Bit. & Rebuf. \\ \hline
        bus 1 & 2659.6 & - & 2188.3 & - & 2480.9 & - & 2839.4 & 5.1 & \textbf{2802.1} & \textbf{-} \\ 
        bus 10 & 1467.0 & - & 1389.4 & 4.6 & 1413.8 & - & 1427.7 & 0.1 & \textbf{1490.4} & \textbf{-} \\ 
        bus 11 & 1327.7 & - & 1195.7 & - & 1264.9 & - & 1336.2 & 0.6 & \textbf{1363.8} & \textbf{-} \\ 
        bus 12 & 769.1 & 4.9 & 721.3 & - & 771.3 & 5.2 & 769.1 & 4.0 & \textbf{793.6} & \textbf{-} \\ 
        bus 13 & 852.1 & 9.7 & \textbf{529.8} & \textbf{-} & 623.4 & 2.6 & 629.8 & 5.5 & 666.0 & 0.7 \\ 
        bus 14 & 1675.5 & 8.3 & 933.0 & - & 1586.2 & 4.6 & 1610.6 & 4.6 & \textbf{1638.3} & \textbf{-} \\ 
        bus 15 & 3919.1 & - & 3497.9 & - & 3598.9 & - & \textbf{4141.5} & \textbf{-} & 3888.3 & - \\ 
        bus 16 & 3950.0 & - & 3652.1 & - & 3857.4 & - & \textbf{4145.7} & \textbf{-} & 3897.9 & - \\ 
        bus 17 & 1943.6 & - & 1844.7 & 2.2 & 1777.7 & - & 2005.3 & - & \textbf{2022.3} & \textbf{-} \\ 
        bus 18 & 1623.4 & - & 1559.6 & 0.4 & 1596.8 & - & 1588.3 & 1.0 & \textbf{1643.6} & \textbf{-} \\ 
        bus 19 & 1472.3 & - & 1212.8 & - & 1461.7 & - & 1440.4 & - & \textbf{1517.0} & \textbf{-} \\ 
        bus 2 & 1820.2 & - & 1656.4 & - & 1737.2 & - & 1861.7 & - & \textbf{1919.1} & \textbf{-} \\ 
        bus 20 & 1726.6 & - & 1517.0 & 1.6 & 1703.2 & - & 1752.1 & - & \textbf{1795.7} & \textbf{-} \\ 
        bus 21 & 1669.1 & - & 1497.9 & - & 1526.6 & - & 1759.6 & 2.1 & \textbf{1730.9} & \textbf{-} \\ 
        bus 22 & \textbf{1743.6} & \textbf{-} & 1594.7 & 7.8 & 1568.1 & - & 1791.5 & 3.6 & 1806.4 & 3.3 \\ 
        bus 23 & 1176.6 & 2.3 & 985.1 & 11.7 & 1076.6 & - & 1195.7 & 4.3 & \textbf{1178.7} & \textbf{-} \\ 
        bus 3 & 1331.9 & - & 1258.5 & - & 1283.0 & - & 1348.9 & 0.9 & \textbf{1368.1} & \textbf{-} \\ 
        bus 4 & 2121.3 & - & 1710.6 & - & 2103.2 & - & 2160.6 & 1.2 & \textbf{2156.4} & \textbf{-} \\ 
        bus 5 & 950.0 & 1.9 & 863.8 & 14.0 & 922.3 & - & \textbf{964.9} & \textbf{-} & 966.0 & - \\ 
        bus 6 & 1568.1 & - & 1324.5 & 0.3 & 1520.2 & - & 1606.4 & - & \textbf{1680.9} & \textbf{-} \\ 
        bus 7 & \textbf{2669.1} & \textbf{-} & 2376.6 & - & 2460.6 & - & 2571.3 & - & 2668.1 & - \\ 
        bus 8 & 2230.9 & - & 2008.5 & 2.9 & 2093.6 & - & \textbf{2305.3} & \textbf{-} & 2290.4 & - \\ 
        bus 9 & 2192.6 & - & 2014.9 & 4.0 & 2079.8 & - & 2297.9 & 1.8 & \textbf{2283.0} & \textbf{-} \\ 
        car 1 & 1440.4 & 0.8 & 1413.8 & 10.1 & 1467.0 & 12.7 & 1447.9 & 4.0 & \textbf{1434.0} & \textbf{-} \\ 
        car 10 & 1300.0 & - & 1127.7 & - & 1218.1 & - & 1338.3 & 2.6 & \textbf{1350.0} & \textbf{-} \\ 
        car 11 & 920.2 & 1.7 & 673.4 & 1.0 & 925.5 & - & 889.4 & - & \textbf{969.1} & \textbf{-} \\ 
        car 12 & 1291.5 & 18.2 & \textbf{1283.0} & \textbf{6.4} & 1291.5 & 13.7 & 1308.5 & 24.9 & 1325.5 & 13.5 \\ 
        car 2 & 1029.8 & - & 903.2 & - & 996.8 & - & 1034.0 & - & \textbf{1060.6} & \textbf{-} \\ 
        car 3 & 1351.1 & 1.8 & 1293.6 & - & 1334.0 & 2.0 & 1309.6 & 0.8 & \textbf{1327.7} & \textbf{-} \\ 
        car 4 & 1433.0 & 1.6 & 1327.7 & 3.4 & 1314.9 & - & 1423.4 & 1.7 & \textbf{1476.6} & \textbf{-} \\ 
        car 5 & 927.7 & 2.1 & 836.2 & 0.1 & 878.7 & - & 895.7 & - & \textbf{959.6} & \textbf{-} \\ 
        car 6 & 1295.7 & - & 1133.0 & - & 1263.8 & - & 1286.2 & - & \textbf{1326.6} & \textbf{-} \\ 
        car 7 & 710.6 & 1.3 & 568.1 & 11.9 & 666.0 & - & 677.7 & 0.9 & \textbf{697.9} & \textbf{-} \\ 
        car 8 & 1688.3 & - & 1448.9 & - & 1676.6 & - & 1704.3 & - & \textbf{1748.9} & \textbf{-} \\ 
        car 9 & 1485.1 & - & 1366.0 & - & 1423.4 & - & 1459.6 & - & \textbf{1526.6} & \textbf{-} \\ 
        ferry 1 & 1454.3 & 0.3 & 1423.4 & - & 1441.5 & - & 1455.3 & - & \textbf{1492.6} & \textbf{-} \\ 
        ferry 10 & 1119.1 & 0.2 & 1021.3 & 0.2 & 1104.3 & 0.2 & 1104.3 & 0.2 & \textbf{1133.0} & \textbf{0.2} \\ 
        ferry 11 & 1038.3 & 1.2 & 847.9 & - & 978.7 & - & 1037.2 & 0.1 & \textbf{1067.0} & \textbf{-} \\ 
        ferry 12 & 1525.5 & - & 1324.5 & 6.4 & 1466.0 & - & 1555.3 & - & \textbf{1586.2} & \textbf{-} \\ 
        ferry 13 & 1421.3 & 0.5 & 1192.6 & 12.6 & \textbf{1320.2} & \textbf{-} & 1436.2 & 5.4 & 1471.3 & 1.8 \\ 
        ferry 14 & 2036.2 & - & 1850.0 & 0.3 & 2074.5 & - & 2098.9 & - & \textbf{2124.5} & \textbf{-} \\ 
        ferry 15 & 1551.1 & - & 1385.1 & - & 1535.1 & - & 1486.2 & 1.1 & \textbf{1573.4} & \textbf{-} \\ 
        ferry 16 & 2684.0 & - & 2406.4 & - & 2480.9 & - & 2734.0 & - & \textbf{2817.0} & \textbf{-} \\ 
        ferry 17 & 859.6 & 0.1 & 768.1 & 9.4 & 818.1 & - & 826.6 & - & \textbf{850.0} & \textbf{-} \\ 
        ferry 18 & 724.5 & 2.8 & 587.2 & 4.9 & 666.0 & 1.1 & 667.0 & 1.0 & \textbf{684.0} & \textbf{-} \\ 
        ferry 19 & 839.4 & 4.8 & 654.3 & - & \textbf{826.6} & \textbf{-} & 825.5 & - & 863.8 & 7.1 \\ 
        ferry 2 & 1012.8 & 3.3 & 951.1 & 10.6 & 990.4 & - & 1001.1 & - & \textbf{1024.5} & \textbf{-} \\ \hline
    \end{tabular}
\end{table*}
\begin{table*}[]
    \centering
    \caption{Details of ABR Tasks}
    \label{detail:2}
    \begin{tabular}{c|c|c|c|c|c|c|c|c|c|c}
    \hline
        Trace Name & \multicolumn{2}{|c|}{BOLA} & \multicolumn{2}{|c|}{Rate-based} & \multicolumn{2}{|c|}{Pensieve} & \multicolumn{2}{|c|}{RobustMPC} & \multicolumn{2}{|c}{Zwei} \\ \hline
        Trace Name & Avg. Bit. & Rebuf. & Avg. Bit. & Rebuf. & Avg. Bit. & Rebuf. & Avg. Bit. & Rebuf. & Avg. Bit. & Rebuf. \\ \hline
        ferry 20 & 686.2 & 6.7 & 635.1 & - & 670.2 & 5.8 & 672.3 & 5.0 & \textbf{674.5} & \textbf{-} \\ 
        ferry 3 & 1211.7 & - & 1093.6 & - & 1170.2 & - & 1231.9 & 0.7 & \textbf{1252.1} & \textbf{-} \\ 
        ferry 4 & 562.8 & 4.2 & 472.3 & 1.9 & 543.6 & - & 558.5 & 4.7 & \textbf{563.8} & \textbf{-} \\ 
        ferry 5 & 1011.7 & 1.7 & 901.1 & 12.0 & 940.4 & - & 1018.1 & 2.2 & \textbf{1027.7} & \textbf{-} \\ 
        ferry 6 & 2014.9 & - & 1598.9 & - & 2039.4 & - & 2093.6 & - & \textbf{2119.1} & \textbf{-} \\ 
        ferry 7 & 1367.0 & 0.5 & 1231.9 & 1.4 & 1309.6 & - & 1346.8 & 0.6 & \textbf{1422.3} & \textbf{-} \\ 
        ferry 8 & 1078.7 & 3.8 & 1010.6 & - & 1076.6 & 1.7 & 1078.7 & 2.5 & \textbf{1105.3} & \textbf{-} \\ 
        ferry 9 & 964.9 & 3.8 & 787.2 & 5.1 & 868.1 & 3.7 & \textbf{898.9} & \textbf{1.9} & 1004.3 & 3.3 \\ 
        metro 1 & 1350.0 & - & 1141.5 & - & 1269.1 & - & 1371.3 & - & \textbf{1389.4} & \textbf{-} \\ 
        metro 10 & 896.8 & - & 625.5 & 3.7 & 837.2 & - & 895.7 & - & \textbf{935.1} & \textbf{-} \\ 
        metro 2 & 1364.9 & - & 1285.1 & 2.8 & 1316.0 & - & 1381.9 & 2.5 & \textbf{1409.6} & \textbf{-} \\ 
        metro 3 & 934.0 & - & 481.9 & - & 884.0 & - & 913.8 & - & \textbf{956.4} & \textbf{-} \\ 
        metro 4 & 1086.2 & 1.0 & 884.0 & 1.2 & 968.1 & 0.6 & 1030.9 & - & \textbf{1121.3} & \textbf{-} \\ 
        metro 5 & 1178.7 & 1.0 & 1041.5 & 3.8 & 1142.6 & - & \textbf{1171.3} & \textbf{-} & 1229.8 & 0.5 \\ 
        metro 6 & 577.7 & 0.9 & 386.2 & 6.6 & 558.5 & - & 548.9 & - & \textbf{573.4} & \textbf{-} \\ 
        metro 7 & 945.7 & - & 769.1 & - & 948.9 & 0.2 & 961.7 & 0.1 & \textbf{977.7} & \textbf{-} \\ 
        metro 8 & 740.4 & 2.5 & 443.6 & 6.4 & 738.3 & - & 724.5 & - & \textbf{762.8} & \textbf{-} \\ 
        metro 9 & 788.3 & - & 702.1 & 4.4 & 735.1 & - & 788.3 & 1.5 & \textbf{806.4} & \textbf{-} \\ 
        train 1 & 1028.7 & 4.2 & 863.8 & 2.3 & 996.8 & - & 1008.5 & - & \textbf{1024.5} & \textbf{-} \\ 
        train 10 & 1838.3 & - & 1627.7 & - & 1709.6 & - & 1894.7 & - & \textbf{1907.4} & \textbf{-} \\ 
        train 11 & 1256.4 & 0.4 & 1163.8 & 0.2 & 1171.3 & - & 1257.4 & - & \textbf{1286.2} & \textbf{-} \\ 
        train 12 & 1304.3 & 0.8 & 911.7 & 1.3 & 1114.9 & - & 1212.8 & - & \textbf{1380.9} & \textbf{-} \\ 
        train 13 & 1560.6 & - & 1236.2 & - & 1334.0 & - & 1564.9 & - & \textbf{1675.5} & \textbf{-} \\ 
        train 14 & 1330.9 & - & 1206.4 & - & 1259.6 & - & 1320.2 & - & \textbf{1356.4} & \textbf{-} \\ 
        train 15 & 1900.0 & - & 1423.4 & - & 1650.0 & - & \textbf{1952.1} & \textbf{-} & 1944.7 & - \\ 
        train 16 & 1661.7 & - & 1475.5 & 3.5 & 1484.0 & - & 1575.5 & - & \textbf{1736.2} & \textbf{-} \\ 
        train 17 & 1620.2 & - & 1394.7 & 4.1 & 1556.4 & - & 1637.2 & - & \textbf{1706.4} & \textbf{-} \\ 
        train 18 & 1978.7 & - & 1725.5 & - & 1803.2 & - & 2012.8 & - & \textbf{2091.5} & \textbf{-} \\ 
        train 19 & 1423.4 & - & 1230.9 & 2.5 & 1302.1 & - & 1452.1 & - & \textbf{1484.0} & \textbf{-} \\ 
        train 2 & 635.1 & 3.5 & 481.9 & 11.1 & \textbf{577.7} & \textbf{-} & 581.9 & - & 664.9 & 3.0 \\ 
        train 20 & 1331.9 & - & 1052.1 & - & 1343.6 & - & 1357.4 & - & \textbf{1373.4} & \textbf{-} \\ 
        train 21 & 1598.9 & - & 1421.3 & - & 1596.8 & - & 1626.6 & - & \textbf{1652.1} & \textbf{-} \\ 
        train 3 & 648.9 & 1.6 & 529.8 & 8.6 & 619.1 & - & 596.8 & - & \textbf{629.8} & \textbf{-} \\ 
        train 4 & 1084.0 & 3.3 & 868.1 & - & 1024.5 & - & 1059.6 & 0.3 & \textbf{1127.7} & \textbf{-} \\ 
        train 5 & 1241.5 & 1.5 & 946.8 & 0.4 & 1155.3 & - & 1151.1 & - & \textbf{1283.0} & \textbf{-} \\ 
        train 6 & 969.1 & 0.9 & 721.3 & - & 875.5 & - & 936.2 & 0.9 & \textbf{991.5} & \textbf{-} \\ 
        train 7 & 825.5 & 8.3 & 750.0 & - & \textbf{784.0} & \textbf{-} & 790.4 & - & 823.4 & 2.6 \\ 
        train 8 & 706.4 & 1.0 & 510.6 & - & 666.0 & - & 693.6 & - & \textbf{730.9} & \textbf{-} \\ 
        train 9 & 855.3 & - & 596.8 & - & 776.6 & - & 852.1 & - & \textbf{881.9} & \textbf{-} \\ 
        tram 1 & 635.1 & - & 472.3 & 1.6 & 625.5 & - & 635.1 & - & \textbf{656.4} & \textbf{-} \\ 
        tram 10 & 869.1 & 2.5 & 673.4 & - & 850.0 & - & \textbf{878.7} & \textbf{-} & 871.3 & - \\ 
        tram 11 & 587.2 & 4.3 & 472.3 & 1.3 & 558.5 & - & 548.9 & - & \textbf{580.9} & \textbf{-} \\ 
        tram 12 & 807.4 & - & 548.9 & - & 795.7 & - & 816.0 & - & \textbf{826.6} & \textbf{-} \\ 
        tram 13 & 788.3 & - & 596.8 & 0.9 & 739.4 & - & 787.2 & - & \textbf{831.9} & \textbf{-} \\ 
        tram 14 & 839.4 & 2.2 & 644.7 & 1.1 & 808.5 & - & 797.9 & - & \textbf{844.7} & \textbf{-} \\ 
        tram 15 & 453.2 & 6.9 & 357.4 & 5.3 & 414.9 & 1.8 & 424.5 & - & \textbf{437.2} & \textbf{-} \\ 
        tram 16 & \textbf{683.0} & \textbf{-} & 548.9 & - & 673.4 & - & 692.6 & - & 692.6 & - \\ 
        tram 17 & \textbf{654.3} & \textbf{-} & 424.5 & - & 625.5 & - & 635.1 & - & 663.8 & - \\ 
        tram 18 & 730.9 & 2.8 & 587.2 & - & 691.5 & - & 706.4 & - & \textbf{751.1} & \textbf{-} \\ \hline
    \end{tabular}
\end{table*}
\begin{table*}[]
    \centering
    \caption{Details of ABR Tasks}
    \label{detail:3}
    \begin{tabular}{c|c|c|c|c|c|c|c|c|c|c}
    \hline
        Trace Name & \multicolumn{2}{|c|}{BOLA} & \multicolumn{2}{|c|}{Rate-based} & \multicolumn{2}{|c|}{Pensieve} & \multicolumn{2}{|c|}{RobustMPC} & \multicolumn{2}{|c}{Zwei} \\ \hline
        Trace Name & Avg. Bit. & Rebuf. & Avg. Bit. & Rebuf. & Avg. Bit. & Rebuf. & Avg. Bit. & Rebuf. & Avg. Bit. & Rebuf. \\ \hline
        tram 19 & 967.0 & 0.1 & 788.3 & - & 907.4 & - & 973.4 & - & \textbf{985.1} & \textbf{-} \\ 
        tram 2 & 577.7 & 3.3 & 434.0 & - & 548.9 & - & 548.9 & - & \textbf{559.6} & \textbf{-} \\ 
        tram 20 & 930.9 & - & 730.9 & - & 885.1 & - & 928.7 & - & \textbf{988.3} & \textbf{-} \\ 
        tram 21 & 692.6 & 0.5 & 558.5 & - & 677.7 & - & 696.8 & - & \textbf{711.7} & \textbf{-} \\ 
        tram 22 & 510.6 & 0.3 & 376.6 & - & 491.5 & - & 501.1 & - & \textbf{506.4} & \textbf{-} \\ 
        tram 23 & 778.7 & - & 529.8 & - & 738.3 & - & 778.7 & - & \textbf{808.5} & \textbf{-} \\ 
        tram 24 & 788.3 & - & 616.0 & 3.4 & 739.4 & - & 769.1 & 0.3 & \textbf{808.5} & \textbf{-} \\ 
        tram 25 & 907.4 & - & 711.7 & - & 813.8 & - & 888.3 & - & \textbf{950.0} & \textbf{-} \\ 
        tram 26 & 568.1 & 0.8 & 300.0 & - & 539.4 & - & \textbf{568.1} & \textbf{-} & 567.0 & - \\ 
        tram 27 & \textbf{692.6} & \textbf{-} & 491.5 & - & 673.4 & - & 692.6 & - & 698.9 & - \\ 
        tram 28 & 558.5 & 0.3 & 386.2 & 2.2 & 510.6 & - & \textbf{539.4} & \textbf{-} & 540.4 & - \\  
        tram 29 & 510.6 & 3.4 & 338.3 & - & 481.9 & - & \textbf{501.1} & \textbf{-} & 495.7 & - \\ 
        tram 3 & 577.7 & 1.3 & 414.9 & - & \textbf{539.4} & \textbf{-} & 534.0 & 1.9 & 570.2 & 1.7 \\ 
        tram 30 & 692.6 & - & 319.1 & - & 644.7 & - & 687.2 & - & \textbf{719.1} & \textbf{-} \\ 
        tram 31 & 769.1 & - & 644.7 & - & 753.2 & - & 769.1 & - & \textbf{783.0} & \textbf{-} \\ 
        tram 32 & \textbf{788.3} & \textbf{-} & 558.5 & 1.7 & 733.0 & - & 750.0 & - & 788.3 & - \\ 
        tram 33 & \textbf{874.5} & \textbf{-} & 644.7 & - & 841.5 & - & 877.7 & 0.2 & 880.9 & - \\ 
        tram 34 & 926.6 & - & 778.7 & 1.6 & 917.0 & - & 930.9 & - & \textbf{959.6} & \textbf{-} \\ 
        tram 35 & 548.9 & 5.0 & 405.3 & 1.6 & 501.1 & - & 486.2 & - & \textbf{550.0} & \textbf{-} \\ 
        tram 36 & 850.0 & 0.3 & 539.4 & - & 819.1 & - & 816.0 & - & \textbf{874.5} & \textbf{-} \\ 
        tram 37 & 984.0 & - & 807.4 & - & 894.7 & - & 1003.2 & - & \textbf{1026.6} & \textbf{-} \\ 
        tram 38 & 644.7 & 9.6 & 481.9 & 4.5 & \textbf{591.5} & \textbf{-} & 625.5 & 4.9 & 646.8 & 2.3 \\ 
        tram 39 & 855.3 & - & 721.3 & - & 814.9 & - & 826.6 & - & \textbf{868.1} & \textbf{-} \\ 
        tram 4 & 462.8 & 5.2 & 300.0 & - & 414.9 & - & 414.9 & - & \textbf{428.7} & \textbf{-} \\ 
        tram 40 & 797.9 & 0.6 & 520.2 & 0.6 & 770.2 & 0.6 & 772.3 & 0.8 & \textbf{827.7} & \textbf{0.6} \\ 
        tram 41 & 769.1 & 2.8 & 663.8 & - & 734.0 & - & 750.0 & 0.5 & \textbf{785.1} & \textbf{-} \\ 
        tram 42 & 1069.1 & 9.0 & 692.6 & - & 961.7 & - & 1057.4 & - & \textbf{1076.6} & \textbf{-} \\ 
        tram 43 & 1079.8 & - & 960.6 & 0.2 & 1048.9 & - & 1084.0 & - & \textbf{1105.3} & \textbf{-} \\ 
        tram 44 & 869.1 & - & 759.6 & - & 801.1 & - & 874.5 & 0.7 & \textbf{893.6} & \textbf{-} \\ 
        tram 45 & 648.9 & 1.1 & 472.3 & 7.0 & \textbf{627.7} & \textbf{0.5} & 635.1 & 0.5 & 627.7 & 0.5 \\ 
        tram 46 & 683.0 & 4.2 & 414.9 & - & 606.4 & - & 662.8 & 0.1 & \textbf{679.8} & \textbf{-} \\ 
        tram 47 & 788.3 & 0.9 & 539.4 & - & 705.3 & - & 764.9 & - & \textbf{795.7} & \textbf{-} \\ 
        tram 48 & 616.0 & 0.3 & 395.7 & - & 558.5 & - & 605.3 & 0.2 & \textbf{619.1} & \textbf{-} \\ 
        tram 49 & 963.8 & 1.1 & 778.7 & - & 963.8 & - & 985.1 & 2.2 & \textbf{1007.4} & \textbf{-} \\ 
        tram 5 & 654.3 & 5.2 & 443.6 & 2.6 & 587.2 & - & 616.0 & - & \textbf{640.4} & \textbf{-} \\ 
        tram 50 & 863.8 & 6.8 & 797.9 & - & 851.1 & - & 845.7 & - & \textbf{864.9} & \textbf{-} \\ 
        tram 51 & 658.5 & 5.5 & 587.2 & - & \textbf{638.3} & \textbf{-} & 647.9 & - & 674.5 & 4.6 \\ 
        tram 52 & 920.2 & 3.4 & 788.3 & 1.7 & 905.3 & - & 903.2 & - & \textbf{927.7} & \textbf{-} \\ 
        tram 53 & 657.4 & 17.3 & 443.6 & 3.8 & 434.0 & 0.2 & 428.7 & - & \textbf{522.3} & \textbf{-} \\ 
        tram 54 & 1123.4 & 6.8 & 862.8 & 6.6 & 1059.6 & - & \textbf{1109.6} & \textbf{-} & 1106.4 & - \\ 
        tram 55 & 1176.6 & 0.4 & 1100.0 & - & 1141.5 & - & 1146.8 & 1.9 & \textbf{1177.7} & \textbf{-} \\ 
        tram 56 & 971.3 & - & 797.9 & - & 944.7 & - & 957.4 & - & \textbf{992.6} & \textbf{-} \\ 
        tram 6 & 740.4 & - & 587.2 & - & 683.0 & - & 744.7 & - & \textbf{784.0} & \textbf{-} \\ 
        tram 7 & 759.6 & - & 596.8 & - & 720.2 & - & 744.7 & - & \textbf{797.9} & \textbf{-} \\ 
        tram 8 & 501.1 & 3.5 & 414.9 & - & 472.3 & - & 462.8 & 0.2 & \textbf{483.0} & \textbf{-} \\ 
        tram 9 & 606.4 & 0.2 & 328.7 & - & 568.1 & - & 572.3 & - & \textbf{629.8} & \textbf{-} \\ \hline
    \end{tabular}
\end{table*}

\end{document}